\documentclass{article}
\usepackage{spconf,amsmath,graphicx}
\usepackage{booktabs}
\usepackage{multirow}
\usepackage[table,xcdraw]{xcolor}
\usepackage{hyperref}
\newcommand{\squeezeup}{\vspace{-1.6mm}}
\title{VocalSound: A Dataset for Improving Human Vocal Sounds Recognition}

\name{Yuan Gong$^1$, Jin Yu$^2$, James Glass$^1$}
\address{$^1$MIT CSAIL, Cambridge, MA 02139, USA \quad
$^2$Signify Research, Cambridge, MA 02142, USA \\
\texttt{\{yuangong,glass\}@mit.edu, jin.yu@signify.com} \squeezeup\squeezeup\squeezeup}

\begin{document}
%\ninept
%
\maketitle
\begin{abstract}

% 3504 total subjects
% 3365 unique subjects
% 21024 total recordings
% 10 percent for 
% country 60
%Counter({' male': 1853, 55.05% ' female': 1499, 44.5% ' other': 13 0.4%})
% [(' United States of America', 0.6035661218424963), (' India', 0.10757800891530461), (' Brazil', 0.08261515601783061), (' Italy', 0.07667161961367014), (' United Kingdom', 0.054680534918276374), (' Canada', 0.022288261515601784), (' Spain', 0.00950965824665676), (' France', 0.008023774145616641), (' Germany', 0.005052005943536404), (' Venezuela', 0.00237741456166419)]

% language 47
%[(' EN', 0.6722139673105498), (' PT', 0.08231797919762258), (' IT', 0.06924219910846954), (' TA', 0.03833580980683507), (' HI', 0.027934621099554236), (' ES', 0.024962852897474), (' ML', 0.015156017830609212), (' TE', 0.010401188707280832), (' FR', 0.009212481426448737), (' DE', 0.004160475482912333)]

% health
%Counter({' no': 3244, ' yes': 121})

% age group 1
% <=25
% age group 2
% <=48
% age group 3
% >48

% 5 and 13 samples in total. 

% questions: if we need to mention irb?
% TODO: figure 2 has a typo 1208 should be 1280

% previous review link: https://docs.google.com/document/d/1D1oGyDeB2dVn0-iHLWqRu3ANahZvZe3NgfTq7QUSr68/edit?usp=sharing
Recognizing human non-speech vocalizations is an important task and has broad applications such as automatic sound transcription and health condition monitoring. However, existing datasets have a relatively small number of vocal sound samples or noisy labels. As a consequence, state-of-the-art audio event classification models may not perform well in detecting human vocal sounds. To support research on building robust and accurate vocal sound recognition, we have created a \emph{VocalSound} dataset consisting of over 21,000 crowdsourced recordings of laughter, sighs, coughs, throat clearing, sneezes, and sniffs from 3,365 unique subjects. Experiments show that the vocal sound recognition performance of a model can be significantly improved by 41.9\% by adding VocalSound dataset to an existing dataset as training material. In addition, different from previous datasets, the VocalSound dataset contains meta information such as speaker age, gender, native language, country, and health condition. Dataset and code available at \href{https://github.com/yuangongnd/vocalsound}{\color{blue}{https://github.com/yuangongnd/vocalsound}}.

\end{abstract}
\section{Introduction}
\squeezeup

Automatic human vocal sound recognition is an important task and has a wide range of applications, e.g., it can help the automatic speech recognition system transcribe both speech and non-speech vocalizations. Recognizing health-related sounds like cough and sneeze could also provide insights into the general well-being of occupants in the office, at home, or other public or private spaces, e.g., the detection of coughing and sneezing and their density, intensity, and other features could be used as an indicator of group health~\cite{kvapilova2019continuous,simou2021universal,larson2011accurate}. 

To build an accurate and robust non-speech vocal sounds recognizer, a dataset with reasonable volume and variety, and accurate annotation is crucial. However, to our knowledge, currently, there is no such large-scale publicly available vocal sound dataset. Moreover, it has been found that a generic audio event classification model trained with existing datasets such as AudioSet~\cite{gemmeke2017audio} does not perform well in classifying human vocal sounds, e.g., the average precision of state-of-the-art models on cough and sneeze classes are only around 0.5 on the AudioSet evaluation set~\cite{kong2020panns,gong2021psla}.  Potential reasons include the fact that corpora such as ESC-50~\cite{piczak2015esc}, FSD50K~\cite{fonseca2020fsd50k}, and AudioSet~\cite{gemmeke2017audio} have a relatively small number of human vocal sound samples (summarized in Table~\ref{tab:vssample}) and the AudioSet annotation quality for these sounds may be lacking due to the challenge of annotating with a large sound vocabulary~\cite{gong2021psla,fonseca2020addressing,shah2018closer}. To address this limitation, in this paper we introduce the \emph{VocalSound} dataset consisting of over 21,000 crowdsourced recordings of laughter, sighs, coughs, throat clearing, sneezes, and sniffs, collected via Amazon Mechanical Turk. The VocalSound dataset is class-balanced, collected from 3,365 speakers from 60 countries, with their ages ranging from 18 to 80. To the best of our knowledge, the VocalSound dataset has the largest number of human vocal sound samples. While one potential limitation of VocalSound dataset is the audio samples are not produced spontaneously but acted by the subjects, our experiments show that the model vocal sound recognition performance on an evaluation set consisting of real vocal sounds can be significantly improved by over 41.9\% by adding VocalSound dataset to existing dataset as training material, demonstrating the usefulness of the VocalSound dataset. In addition, in contrast to previous datatsets~\cite{piczak2015esc,fonseca2020fsd50k,gemmeke2017audio}, the VocalSound dataset contains meta information such as speaker age, gender, native language, country, and health condition to support research.

%In our experiments, we find the vocal sound recognition accuracy differs for different age and gender groups. Therefore, the VocalSound dataset could also facilitate future research on decreasing the discrepancy of vocal sound recognition accuracy among different subgroups of people.

\begin{figure*}[t]
\minipage{0.28\textwidth}
  \includegraphics[width=\linewidth]{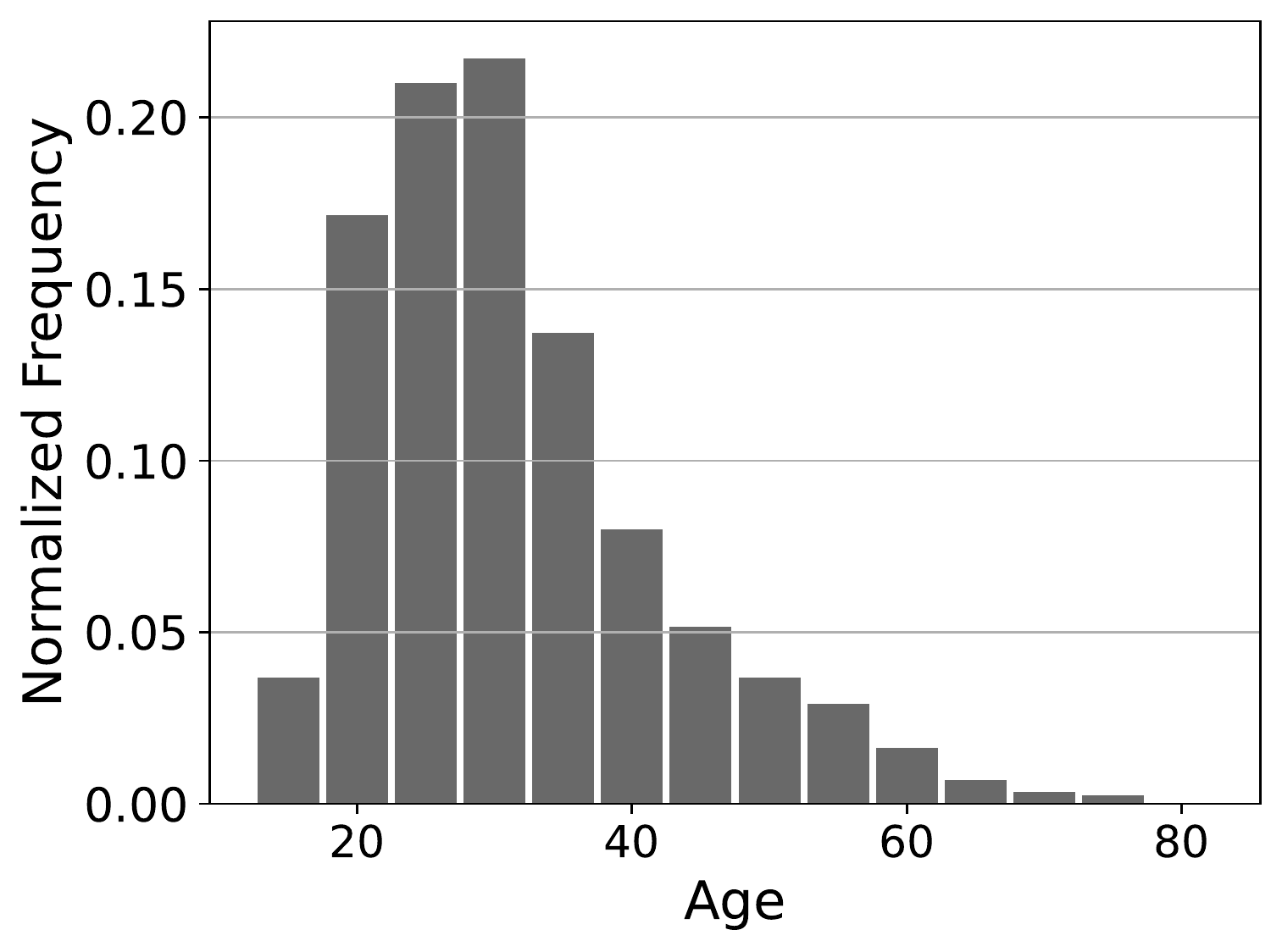}
\endminipage\hfill
\minipage{0.28\textwidth}
  \includegraphics[width=\linewidth]{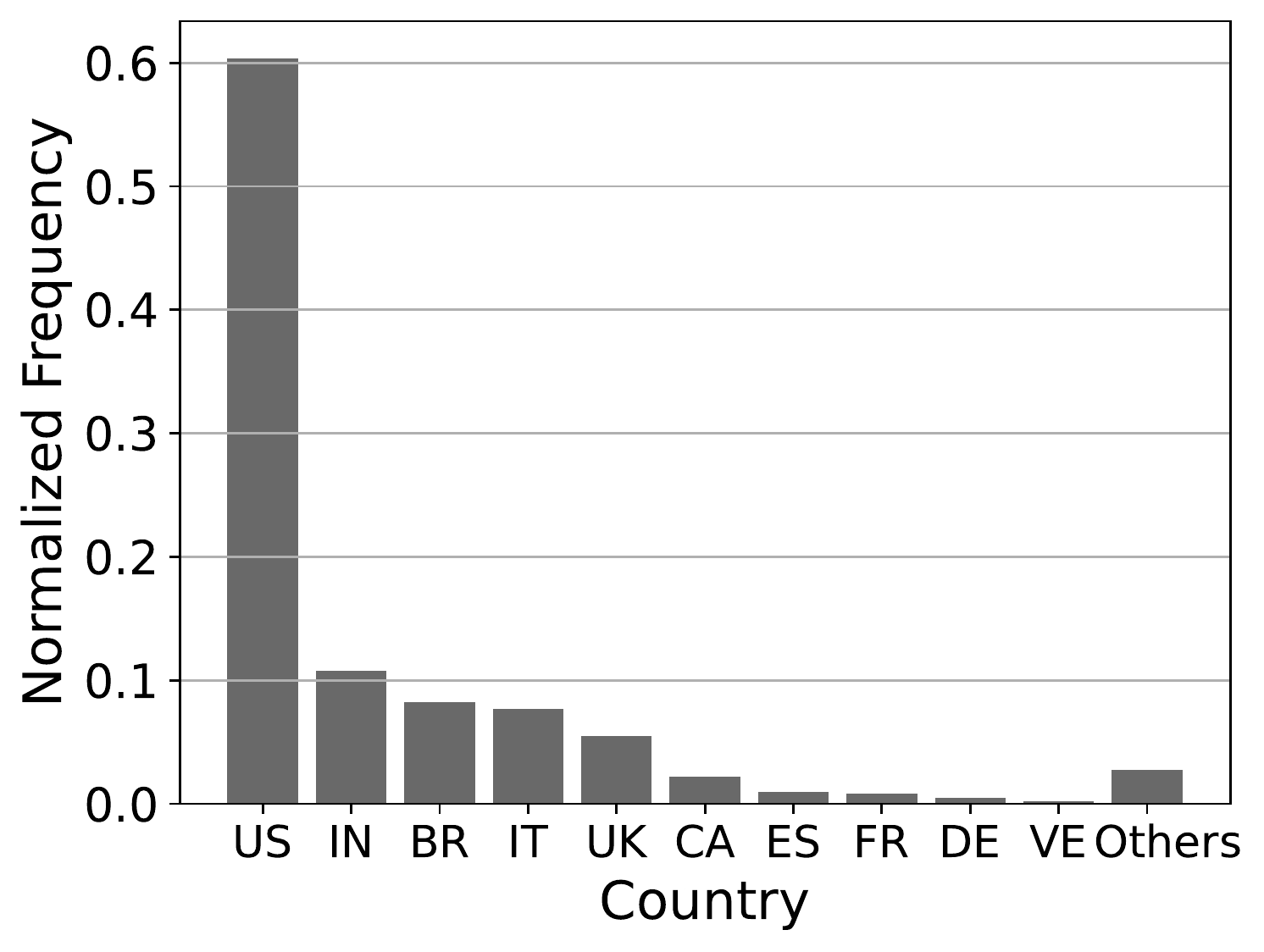}
\endminipage\hfill
\minipage{0.28\textwidth}%
  \includegraphics[width=\linewidth]{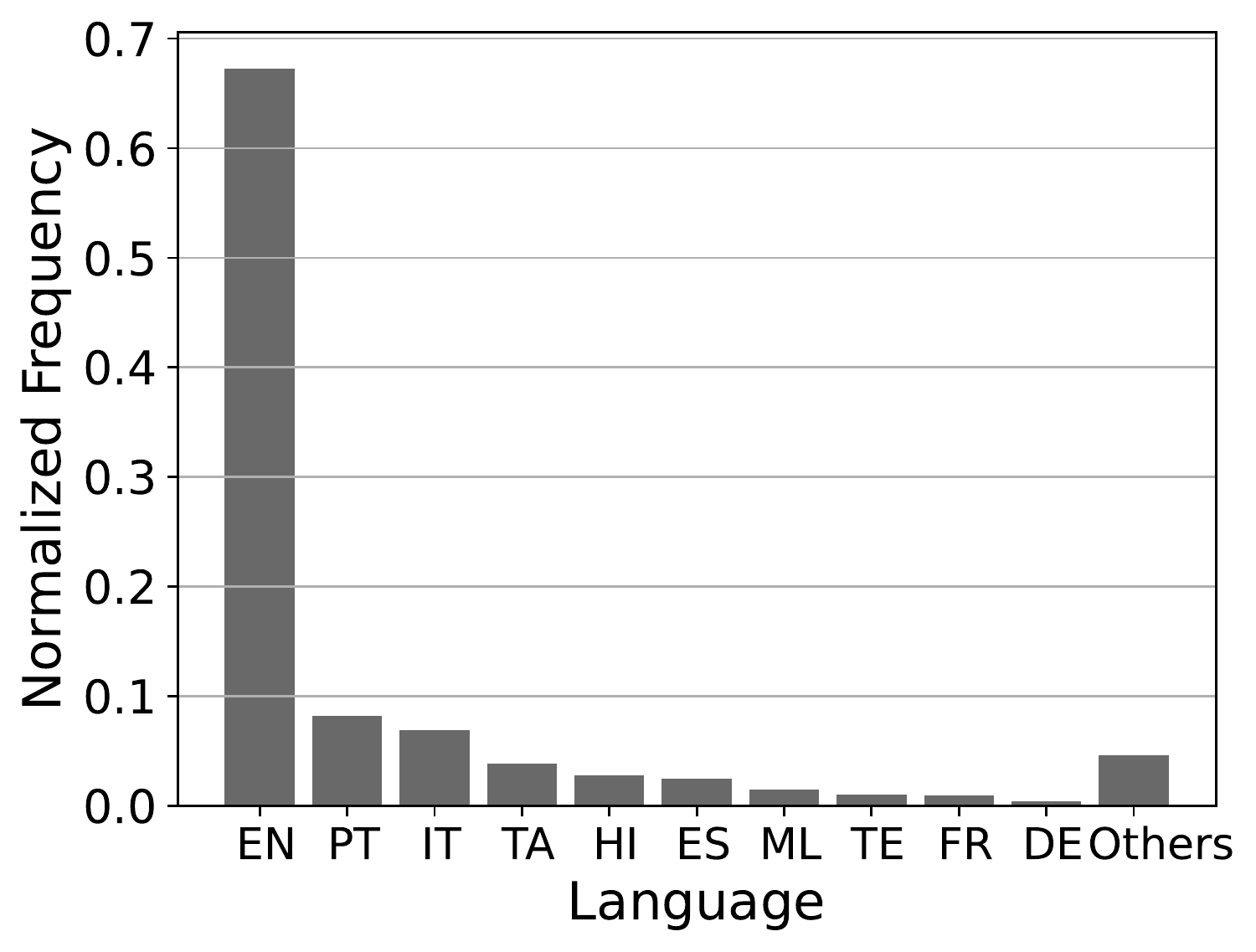}
\endminipage
\caption{The speaker age (left), country (center), and native language (right) distribution.}
\label{fig:dist}
\squeezeup\squeezeup
\end{figure*}

\squeezeup\squeezeup
\section{Related Work}
\squeezeup
\label{sec:related}

There are a few existing datasets for generic audio event classification that also contain human vocal sound samples such as AudioSet~\cite{gemmeke2017audio}, FSD50K~\cite{fonseca2020fsd50k}, ESC-50~\cite{piczak2015esc}, and DCASE~\cite{mesaros2017dcase}. Specifically, AudioSet is currently the largest publicly available dataset for generic audio classification consisting of over 2 million audio clips excised from YouTube and labeled with a set of 527 labels. The FSD50K dataset consists of 51,197 audio clips unequally distributed in 200 sound classes. While AudioSet and FSD50K consist of a large number of audio samples, they are class imbalanced and have a relatively small number of vocal sound samples (summarized in Table~\ref{tab:vssample}). Also, limited by the data acquisition scheme, they do not provide speaker information such as age, gender, native language, etc. Due to the difficulty of annotating YouTube videos with a large sound vocabulary, the noisy label issue has been found in AudioSet~\cite{gong2021psla,fonseca2020addressing,shah2018closer}, which could also impact the performance of the model trained on it. Recently, there are a few efforts to collect cough samples for building COVID-19 classification models~\cite{orlandic2020coughvid,laguarta2020covid,brown2020exploring,cohen2020novel,imran2020ai4covid,schuller2021interspeech,bagad2020cough}. In comparison with the proposed dataset, existing imitated vocal sound datasets~\cite{kim2018vocal,cartwright2015vocalsketch} are much smaller in size.

The proposed VocalSound dataset differs from previous efforts in that 1) the VocalSound dataset is class-balanced and has more vocal sound samples collected from a large number of speakers with reasonable gender and age distributions. Due to the data acquisition scheme, the labels are also reliable. 2) the VocalSound dataset has rich meta information, including speaker gender, age, native language, country, and health condition, which broadens the application of the dataset, e.g., the metadata can be used to study the impact of gender, age, language on the performance of vocal sound classification models; the health label can potentially be used to build speech-based health classification system; the anonymous speaker label can potentially be used to build vocal sound-based speaker re-identification system, etc.

\begin{table}[t]
\setlength\tabcolsep{2.5pt}
\small
\centering
\begin{tabular}{lcccc}
\hline
                              & ESC-50                       & FSD50K                      & AudioSet                  & VocalSound               \\ \hline
Laughter                      & 40                           & 1,186                        & 5,696                     & 3,504                    \\
Sigh                          & -                            & 136                            & 301                       & 3,504                    \\
Cough                         & 40                           & 385                          & 871                       & 3,504                    \\
Throat Clearing               & -                            & -                            & 355                       & 3,504                    \\
Sneeze                        & 40                           & 125                          & 1,200                     & 3,504                    \\
Sniff                         & -                            & -                          & 205                       & 3,504                    \\
{\color[HTML]{656565} Others} & {\color[HTML]{656565} 1,880} & {\color[HTML]{656565} 49.4K} & {\color[HTML]{656565} 2M} & {\color[HTML]{656565} 0} \\ \hline
Vocal Sound Total             & 120                          & 1,832                        & 8,628                     & 21,024                   \\ \hline
\end{tabular}
\caption{Data volume comparison of VocalSound and existing datasets.}
\label{tab:vssample}
\squeezeup
\end{table}

% \begin{figure}[t]
%   \centering
%   \includegraphics[width=6.0cm]{fig/volume_bw.pdf}
%   \caption{The box plot of the audio volume of each sound class.}
%   \label{fig:length}
% \end{figure}

\squeezeup\squeezeup
\section{VocalSound Data Collection}
\squeezeup

We crowdsource the VocalSound recordings via Amazon Mechanical Turk (AMT). Subjects volunteer to complete our Human Intelligence Tasks (HITs) on AMT and get compensation after the HITs are reviewed and approved by us. Our HIT consists of seven subtasks. First, we ask the gender, age, country, native language, and health information of the subject. For the health condition, we ask the question ``do you have a cold, allergy, or other health-related symptoms that might affect your speech today?''. Then in subtasks 2-7, we ask the subject to record themselves laughing, sighing, coughing, clearing their throat, sneezing, and sniffing.  We do not collect personally identifiable information from the subject or the recording device, and the data collection is anonymous. We approve HITs according to the following three criteria: 1) the audio length is longer than 2 seconds; 2) we use Google Speech API to transcribe the audio to make sure no speech is contained, audios that can be transcribed as words (e.g., haha) are manually verified; 3) We use the model in~\cite{gong2021psla} to verify if the audio matches the corresponding class. As the prediction of the model might not be accurate, we only use a low threshold to exclude obvious unrelated samples. We apply these criteria during the data collection process, to provide immediate feedback to the Turker and to improve the overall quality of the recordings. We manually verified 600 samples from the dataset, with about 96\% judged to be high quality recordings.

%In our data collection process there are a small number of samples that fall below the criteria are approved in the early stage of the data collection.  The data collection is approved by the Committee on the Use of Humans as Experimental Subjects of the Massachusetts Institute of Technology.

\squeezeup\squeezeup
\section{Data Distribution}
\label{sec:split}
\squeezeup

We collected 3,504 HITs completed by 3,365 unique subjects. Only a small number of subjects completed the HIT more than one time. Our goal was to encourage as much diversity across speakers as possible.  Among the subjects, 45\% are female, 55\% are male. Therefore, the VocalSound dataset is roughly gender-balanced. We show the subject age, country, and native language distribution in Figure~\ref{fig:dist}. The age of the subjects ranges from 18 to 80 while most subjects' ages fall between 20 to 40. Nevertheless, there are still 321 subjects that are older than 50, which are adequate for evaluating the model performance on the senior group. The subjects are from 60 countries, where the United States (60.3\%), India (10.8\%), and Brazil (8.3\%) are the majority countries. English (67.2\%), Portuguese (8.7\%), and Italian (6.8\%) are the corresponding dominant native languages of the subjects. 4.0\% of the subjects report that they have health-related symptoms that might affect their speech during the data collection. The mean, median, and standard deviation of the audio length is 4.18s, 3.72s, and 1.81s, respectively. The audios are recorded at 44.1kHz in .wav format.

% after revision 15570 tr data, 1860 val data, 3594 te data
% after revision, 18-25: 892 samples, 26-48: 2351 samples, 49-80: 348
% after revision, male: 1869, female: 1704
The data is split into training, validation, and evaluation sets with 15570 (74\%), 1860 (9\%), and 3594 (17\%) samples, respectively. The three sets are speaker-independent. We pay special attention to the evaluation set and manually checked one sample from each speaker and removed low-quality recordings. This \emph{clean} evaluation set makes the model evaluation fairer and more effective.

\begin{figure}[t]
  \centering
  \includegraphics[width=4.65cm]{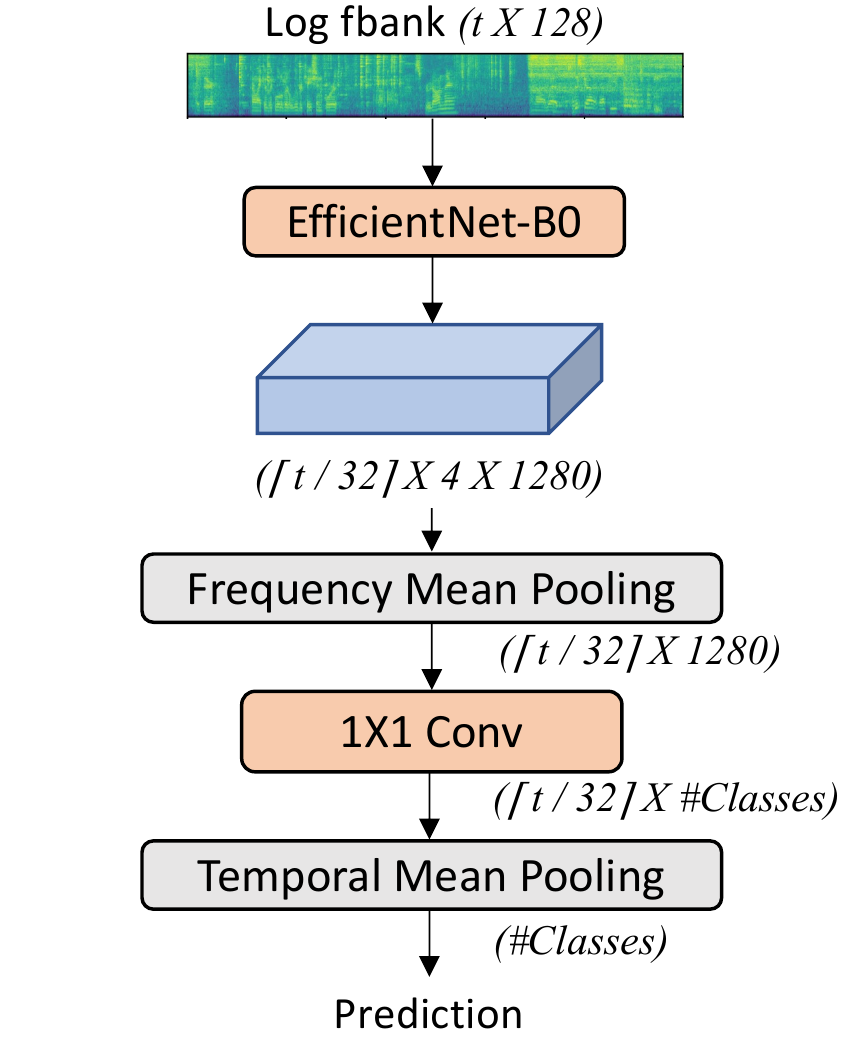}
  \caption{The model architecture used in our experiments. Each audio waveform is first converted to a sequence of 128-dimensional log Mel filterbank (fbank) features computed with a 25ms Hanning window every 10ms. The $t\times128$ fbank feature vector is input to an EfficientNet-B0 model~\cite{tan2019efficientnet}. The EfficientNet-B0 model effectively downsamples the time and frequency dimensions by a factor of 32, and the feature dimension is 1280. Thus, the penultimate output of the model is a $\lceil t/32\rceil\times4\times1280$ tensor. We apply mean pooling over the 4 frequency dimensions to produce a $33\times1408$ representation that is fed to a set of $1\times1$ convolutional filters with a sigmoid activation function, where \#class is the number of prediction classes. A temporal mean pooling is then performed to produce a final \#class dimensional output for each class label.}
  \label{fig:arc}
  \squeezeup\squeezeup
\end{figure}

\squeezeup\squeezeup
\section{Baseline Experiments}
\squeezeup
\label{sec:exp}

We conduct two baseline experiments using the proposed VocalSound dataset. First, in Section~\ref{sec:exp1}, we conduct a six-class (laughter, sigh, cough, throat clearing, sneeze, and sniff) classification experiment on the VocalSound dataset to show the model trained with VocalSound dataset can perform well on vocal sound classification. Second, in Section~\ref{sec:exp2}, we show the VocalSound dataset can help improve the vocal sound recognition from a wide variety of background sounds by combining it with the existing FSD50K dataset.

For both experiments, we use an EfficientNet-B0~\cite{tan2019efficientnet} based audio classifier (illustrated in Figure~\ref{fig:arc}), which has a similar architecture with the state-of-the-art audio classification model in~\cite{gong2021psla}, but uses EfficientNet-B0 and mean temporal pooling instead of EfficientNet-B2 and attention pooling. As discussed in~\cite{gong2021psla}, such simplification can greatly improve the computational efficiency while only marginally reducing performance. For both experiments, we train the model using an Adam optimizer~\cite{kingma2014adam}, an initial learning rate of 1e-4, a batch size of 100, and cross-entropy loss for 50 epochs and select the best model using the development set and evaluate the model on the evaluation set. SpecAugment~\cite{park19e_interspeech} is used during training. We repeat each experiment 3 times and report the mean and standard deviation of the results.

\begin{table}[]
\centering
\small
\begin{tabular}{@{}lc@{}}
\toprule
Test Set                                      & Accuracy (\%)                    \\ \midrule
VocalSound Validation Set                     & 90.1$\pm$0.2                       \\
VocalSound Evaluation Set                     & 90.5$\pm$0.2                       \\ \midrule
\multicolumn{2}{l}{\textit{Different Age Group}}    \\ \midrule
Age 18-25                                     & 91.5$\pm$0.3                       \\
Age 26-48                                     & 90.1$\pm$0.2                       \\
Age 49-80                                     & 90.9$\pm$1.6                       \\ \midrule
\multicolumn{2}{l}{\textit{Different Gender Group}} \\ \midrule
Male                                          & 89.2$\pm$0.5                       \\
Female                                        & 91.9$\pm$0.1                       \\ \bottomrule
\end{tabular}
\caption{Six-class Vocal Sound Classification Results.}
\label{tab:exp1}
\squeezeup\squeezeup
\end{table}

% \label{exp:exp1}
% \begin{figure}[]
%   \centering
%   \includegraphics[width=6cm]{fig/cm_exp1.pdf}
%   \caption{The confusion matrix of the model trained with VocalSound training set on VocalSound evaluation set.}
%   \label{fig:conf_balas}
% \end{figure}
%\begin{tabular}[c]{@{}c@{}}Vocal Classes\\  Overall\end{tabular}
\begin{table*}[t]
\centering
\small
\begin{tabular}{@{}ccccccccccccc@{}}
\toprule
                                            Training Set                  & \multicolumn{2}{c}{Laughter}                                                                                                         & \multicolumn{2}{c}{Sigh}                                                                                                             & \multicolumn{2}{c}{Cough}                                                                                                            & \multicolumn{2}{c}{Sneeze}                                                                                                           & \multicolumn{2}{c}{Background}                                                                                                       & \multicolumn{2}{c}{Vocal Classes Overall}                            \\ \cmidrule(l){2-13} 
                                                              & F1                                                     & \multicolumn{1}{c|}{AP}                                                     & F1                                                     & \multicolumn{1}{c|}{AP}                                                     & F1                                                     & \multicolumn{1}{c|}{AP}                                                     & F1                                                     & \multicolumn{1}{c|}{AP}                                                     & F1                                                     & \multicolumn{1}{c|}{AP}                                                     & Avg. F1                                                & mAP                                                    \\ \midrule
\begin{tabular}[c]{@{}c@{}}FSD50K \\ Only\end{tabular}        & \begin{tabular}[c]{@{}c@{}}0.45\\ $\pm$0.04\end{tabular} & \multicolumn{1}{c|}{\begin{tabular}[c]{@{}c@{}}0.46\\ $\pm$0.05\end{tabular}} & \begin{tabular}[c]{@{}c@{}}0.31\\ $\pm$0.01\end{tabular} & \multicolumn{1}{c|}{\begin{tabular}[c]{@{}c@{}}0.28\\ $\pm$0.02\end{tabular}} & \begin{tabular}[c]{@{}c@{}}0.41\\ $\pm$0.04\end{tabular} & \multicolumn{1}{c|}{\begin{tabular}[c]{@{}c@{}}0.35\\ $\pm$0.02\end{tabular}} & \begin{tabular}[c]{@{}c@{}}0.61\\ $\pm$0.02\end{tabular} & \multicolumn{1}{c|}{\begin{tabular}[c]{@{}c@{}}0.57\\ $\pm$0.07\end{tabular}} & \begin{tabular}[c]{@{}c@{}}0.97\\ $\pm$0.00\end{tabular} & \multicolumn{1}{c|}{\begin{tabular}[c]{@{}c@{}}0.99\\ $\pm$0.00\end{tabular}} & \begin{tabular}[c]{@{}c@{}}0.45\\ $\pm$0.02\end{tabular} & \begin{tabular}[c]{@{}c@{}}0.41\\ $\pm$0.01\end{tabular} \\ \midrule
\begin{tabular}[c]{@{}c@{}}FSD50K+ \\VocalSound\end{tabular} & \begin{tabular}[c]{@{}c@{}}0.59\\ $\pm$0.01\end{tabular} & \multicolumn{1}{c|}{\begin{tabular}[c]{@{}c@{}}0.54\\ $\pm$0.02\end{tabular}} & \begin{tabular}[c]{@{}c@{}}0.41\\ $\pm$0.03\end{tabular} & \multicolumn{1}{c|}{\begin{tabular}[c]{@{}c@{}}0.37\\ $\pm$0.05\end{tabular}} & \begin{tabular}[c]{@{}c@{}}0.65\\ $\pm$0.01\end{tabular} & \multicolumn{1}{c|}{\begin{tabular}[c]{@{}c@{}}0.67\\ $\pm$0.01\end{tabular}} & \begin{tabular}[c]{@{}c@{}}0.71\\ $\pm$0.07\end{tabular} & \multicolumn{1}{c|}{\begin{tabular}[c]{@{}c@{}}0.77\\ $\pm$0.01\end{tabular}} & \begin{tabular}[c]{@{}c@{}}0.98\\ $\pm$0.00\end{tabular} & \multicolumn{1}{c|}{\begin{tabular}[c]{@{}c@{}}0.99\\ $\pm$0.00\end{tabular}} & \begin{tabular}[c]{@{}c@{}}0.59\\ $\pm$0.02\end{tabular} & \begin{tabular}[c]{@{}c@{}}0.59\\ $\pm$0.01\end{tabular} \\ \midrule
Improvement                                                 & 29.7\%                                                 & \multicolumn{1}{c|}{18.1\%}                                                 & 30.5\%                                                 & \multicolumn{1}{c|}{32.2\%}                                                 & 58.6\%                                                 & \multicolumn{1}{c|}{93.9\%}                                                 & 16.0\%                                                 & \multicolumn{1}{c|}{34.3\%}                                                 & 1.5\%                                                  & \multicolumn{1}{c|}{0.0\%}                                                  & 31.8\%                                                 & 41.9\%                                                 \\ \bottomrule
\end{tabular}
\caption{Vocal Sound Recognition Results on FSD50K Evaluation Set.}
\label{tab:exp2}
\squeezeup
\end{table*}

\subsection{Six-class Vocal Sound Classification}
\label{sec:exp1}

In this experiment, we train a six-class (laughter, sigh, cough, throat clearing, sneeze, and sniff) classifier using the VocalSound dataset. We train, validate, and evaluate the model using the training, validation, and evaluation sets mentioned in Section~\ref{sec:split}. We downsample the sampling rate to 16kHz, and truncate or pad all audio samples to 5 seconds. 

As shown in Table~\ref{tab:exp1}, the accuracy on the evaluation set is 90.5$\pm$0.2\% (on the validation set: 90.1$\pm$0.2\%), demonstrating the proposed VocalSound dataset can be used as training material to effectively train a vocal sound classifier. Interestingly, we find the classification accuracy varies with the speaker groups. As shown in Table~\ref{tab:exp1}, the model achieves an accuracy of 91.5$\pm$0.3\%, 90.1$\pm$0.2\%, 90.9$\pm$1.6\% on the age group of 18-25, 26-48, 49-80, respectively; and 89.2$\pm$0.5\% and 91.9$\pm$0.1\% on male and female subjects, respectively. The performance does not solely depend on the number of training samples of each group as the age group of 26-48 and the male group have the largest number of samples but do not have the highest accuracy. Since the VocalSound dataset contains speaker meta information, it can be used to support future research on removing such model bias.

\label{exp:exp1}
\begin{figure}[]
  \centering
  \includegraphics[width=8.15cm]{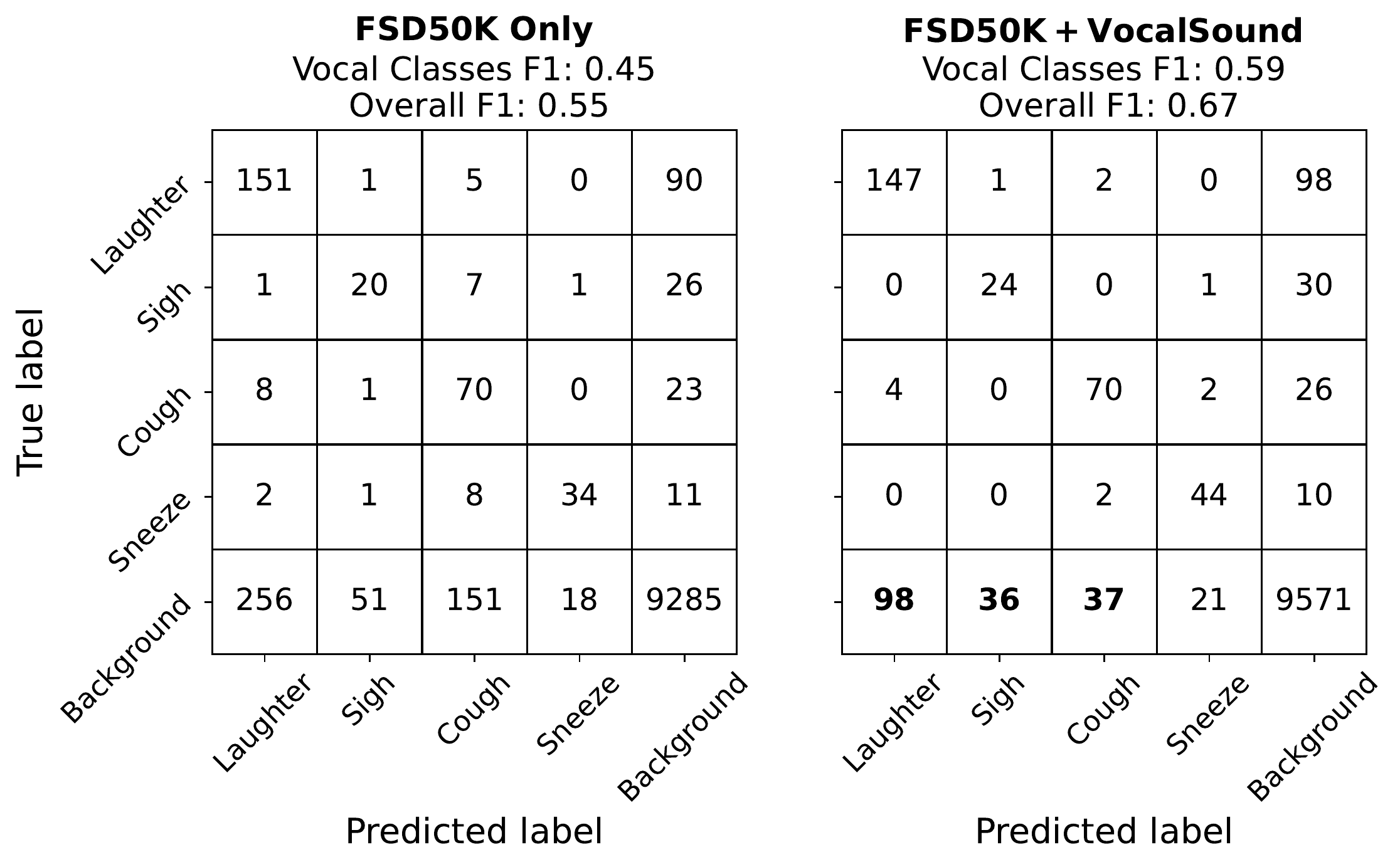}
  \caption{Comparison of the confusion matrixes of the model trained with only FSD50K (left) and with FSD50K + VocalSound (right), evaluated on FSD50K evaluation set. Results averaged from three runs and rounded to integer. Adding VocalSound in the training set can improve the precision of the vocal sound classes (highlighted in bold numbers).}
  \label{fig:exp2}
  \squeezeup\squeezeup
\end{figure}

\squeezeup
\subsection{Vocal Sound Recognition from Background Sounds}
\label{sec:exp2}

While the model trained with just the VocalSound dataset achieves good accuracy on the 6-class vocal sound classification task, recognizing vocal sounds from a wide variety of background natural sounds is a more important and challenging task. Even the state-of-the-art audio classification models in~\cite{gong2021psla,kong2020panns} cannot achieve satisfactory results for the vocal sound classes, e.g., the average precision on cough and sneeze classes are only around 0.5 on the AudioSet evaluation set. In this experiment, we show how the proposed VocalSound dataset can help improve the performance for this task. Specifically, we show that combining the VocalSound dataset with the existing FSD50K dataset as training material can noticeably improve vocal sounds recognition from background sounds compared with only using FSD50K as training material. The reason why we use FSD50K rather than AudioSet as the base dataset is because FSD50K, especially its evaluation set, has more accurate labels~\cite{fonseca2020fsd50k} while labels of AudioSet are relatively noisy~\cite{gong2021psla,fonseca2020addressing,shah2018closer}. The FSD50K consists of 51K audio clips distributed in 200 sound classes so a wide variety of background sounds are included. Since FSD50K only contains 4 vocal sound classes, we consider a 4+1-class (laughter, sigh, cough, and sneeze + background class) classification problem. For the FSD50K dataset, we relabel all samples that are not labeled as laughter, sigh, cough, and sneeze as a new ``background'' class. FSD50K is a multi-label dataset but there are only 5, 1, and 13 samples having more than one vocal sound label in the training, validation, and evaluation set, respectively. We randomly select one label for these samples, making the task a single-class classification problem. 

We compare two training set settings: 1) FSD50K only, we use the official 37k training split of FSD50K as the training set, among the 37k samples, only 1,241 samples are vocal sound samples and other samples are background sounds; 2) FSD50K + VocalSound, VocalSound dataset samples are combined with the FSD50K training set to form a new training set. It is worth mentioning that both datasets are severely class-imbalanced. The background class has 10$\times$ more samples than each vocal sound class even after VocalSound samples are added. Therefore, we use a balanced sampling strategy~\cite{gong2021psla} to make the model see roughly the same number of samples of each class during training, specifically, we use the \texttt{torch.utils.data.WeightedRandomSampler} function. This also makes the comparison between the models trained with these two training sets fairer. In addition to balanced sampling, we also apply SpecAugment~\cite{park19e_interspeech} and random time shift to alleviate the class-imbalance issue.

We train the EfficientNet models with the aforementioned two training sets with the same setting, validate the models using the FSD50K validation set, and evaluate the models using the FSD50K evaluation set. Note that we intentionally evaluate on FSD50K (real sounds, independent from the VocalSound dataset) rather than the VocalSound dataset itself to more fairly show the advantage of adding VocalSound for training. Since the evaluation set is also class-imbalanced, we report average precision (AP) and f1-score (F1) rather than accuracy. As shown in Table~\ref{tab:exp2}, training with FSD50K + VocalSound can significantly boost the vocal sound recognition performance by a relative f1-score improvement of 31.8\% and an average precision improvement of 41.9\%. In Figure~\ref{fig:exp2}, we compare the confusion matrix of the two models. We find that adding VocalSound in the training set can greatly improve the precision of the vocal sound classes. We run a McNemar's test and confirm the improvement is statistically significant ($p<0.05$). All these demonstrate that the proposed VocalSound dataset, while consisting of non-spontaneous sounds, can be used as training material to effectively improve the vocal sound classification performance in realistic use cases.

\squeezeup\squeezeup
\section{Conclusions}
\squeezeup

In this paper, we introduce VocalSound, a new dataset consisting of over 21,000 audio recordings of laughter, sighs, coughs, throat clearing, sneezes, and sniffs. Compared with existing generic audio event datasets, the proposed dataset has more vocal sound samples and richer speaker information. Our experiments show that the VocalSound dataset can noticeably improve vocal sound recognition performance. We hope the new dataset can contribute to future research on building accurate and robust vocal sound recognizers.

\squeezeup\squeezeup
\section{Acknowledgements}
\squeezeup\squeezeup

This work is supported in part by Signify.

%\newpage
\bibliographystyle{IEEEtran}
\bibliography{refs}

\begin{thebibliography}{10}
\providecommand{\url}[1]{#1}
\csname url@samestyle\endcsname
\providecommand{\newblock}{\relax}
\providecommand{\bibinfo}[2]{#2}
\providecommand{\BIBentrySTDinterwordspacing}{\spaceskip=0pt\relax}
\providecommand{\BIBentryALTinterwordstretchfactor}{4}
\providecommand{\BIBentryALTinterwordspacing}{\spaceskip=\fontdimen2\font plus
\BIBentryALTinterwordstretchfactor\fontdimen3\font minus
  \fontdimen4\font\relax}
\providecommand{\BIBforeignlanguage}[2]{{%
\expandafter\ifx\csname l@#1\endcsname\relax
\typeout{** WARNING: IEEEtran.bst: No hyphenation pattern has been}%
\typeout{** loaded for the language `#1'. Using the pattern for}%
\typeout{** the default language instead.}%
\else
\language=\csname l@#1\endcsname
\fi
#2}}
\providecommand{\BIBdecl}{\relax}
\BIBdecl

\bibitem{kvapilova2019continuous}
L.~Kvapilova, V.~Boza, P.~Dubec, M.~Majernik, J.~Bogar, J.~Jamison, J.~C.
  Goldsack, D.~J. Kimmel, and D.~R. Karlin, ``Continuous sound collection using
  smartphones and machine learning to measure cough,'' \emph{Digital
  Biomarkers}, 2019.

\bibitem{simou2021universal}
N.~Simou, N.~Stefanakis, and P.~Zervas, ``A universal system for cough
  detection in domestic acoustic environments,'' in \emph{European Signal
  Processing Conference}, 2021.

\bibitem{larson2011accurate}
E.~C. Larson, T.~Lee, S.~Liu, M.~Rosenfeld, and S.~N. Patel, ``Accurate and
  privacy preserving cough sensing using a low-cost microphone,'' in
  \emph{International Conference on Ubiquitous Computing}, 2011.

\bibitem{gemmeke2017audio}
J.~F. Gemmeke, D.~P. Ellis, D.~Freedman, A.~Jansen, W.~Lawrence, R.~C. Moore,
  M.~Plakal, and M.~Ritter, ``{Audio Set}: An ontology and human-labeled
  dataset for audio events,'' in \emph{ICASSP}, 2017.

\bibitem{kong2020panns}
Q.~Kong, Y.~Cao, T.~Iqbal, Y.~Wang, W.~Wang, and M.~D. Plumbley, ``Panns:
  Large-scale pretrained audio neural networks for audio pattern recognition,''
  \emph{IEEE/ACM Transactions on Audio, Speech, and Language Processing}, 2020.

\bibitem{gong2021psla}
Y.~Gong, Y.-A. Chung, and J.~Glass, ``Psla: Improving audio tagging with
  pretraining, sampling, labeling, and aggregation,'' \emph{IEEE/ACM
  Transactions on Audio, Speech, and Language Processing}, 2021.

\bibitem{piczak2015esc}
K.~J. Piczak, ``{ESC}: Dataset for environmental sound classification,'' in
  \emph{ACM Multimedia}, 2015.

\bibitem{fonseca2020fsd50k}
E.~Fonseca, X.~Favory, J.~Pons, F.~Font, and X.~Serra, ``Fsd50k: an open
  dataset of human-labeled sound events,'' \emph{arXiv preprint
  arXiv:2010.00475}, 2020.

\bibitem{fonseca2020addressing}
E.~Fonseca, S.~Hershey, M.~Plakal, D.~P. Ellis, A.~Jansen, and R.~C. Moore,
  ``Addressing missing labels in large-scale sound event recognition using a
  teacher-student framework with loss masking,'' \emph{IEEE Signal Processing
  Letters}, 2020.

\bibitem{shah2018closer}
A.~Shah, A.~Kumar, A.~G. Hauptmann, and B.~Raj, ``A closer look at weak label
  learning for audio events,'' \emph{arXiv preprint arXiv:1804.09288}, 2018.

\bibitem{mesaros2017dcase}
A.~Mesaros, T.~Heittola, A.~Diment, B.~Elizalde, A.~Shah, E.~Vincent, B.~Raj,
  and T.~Virtanen, ``Dcase 2017 challenge setup: Tasks, datasets and baseline
  system,'' in \emph{DCASE}, 2017.

\bibitem{orlandic2020coughvid}
L.~Orlandic, T.~Teijeiro, and D.~Atienza, ``The coughvid crowdsourcing dataset:
  A corpus for the study of large-scale cough analysis algorithms,''
  \emph{arXiv preprint arXiv:2009.11644}, 2020.

\bibitem{laguarta2020covid}
J.~Laguarta, F.~Hueto, and B.~Subirana, ``Covid-19 artificial intelligence
  diagnosis using only cough recordings,'' \emph{IEEE Open Journal of
  Engineering in Medicine and Biology}, 2020.

\bibitem{brown2020exploring}
C.~Brown, J.~Chauhan, A.~Grammenos, J.~Han, A.~Hasthanasombat, D.~Spathis,
  T.~Xia, P.~Cicuta, and C.~Mascolo, ``Exploring automatic diagnosis of
  covid-19 from crowdsourced respiratory sound data,'' in \emph{ACM KDD}, 2020.

\bibitem{cohen2020novel}
M.~Cohen-McFarlane, R.~Goubran, and F.~Knoefel, ``Novel coronavirus cough
  database: Nococoda,'' \emph{IEEE Access}, 2020.

\bibitem{imran2020ai4covid}
A.~Imran, I.~Posokhova, H.~N. Qureshi, U.~Masood, M.~S. Riaz, K.~Ali, C.~N.
  John, M.~I. Hussain, and M.~Nabeel, ``Ai4covid-19: Ai enabled preliminary
  diagnosis for covid-19 from cough samples via an app,'' \emph{Informatics in
  Medicine Unlocked}, 2020.

\bibitem{schuller2021interspeech}
B.~W. Schuller, A.~Batliner, C.~Bergler, C.~Mascolo, J.~Han, I.~Lefter,
  H.~Kaya, S.~Amiriparian, A.~Baird, L.~Stappen \emph{et~al.}, ``The
  interspeech 2021 computational paralinguistics challenge: Covid-19 cough,
  covid-19 speech, escalation \& primates,'' in \emph{Interspeech}, 2021.

\bibitem{bagad2020cough}
P.~Bagad, A.~Dalmia, J.~Doshi, A.~Nagrani, P.~Bhamare, A.~Mahale, S.~Rane,
  N.~Agarwal, and R.~Panicker, ``Cough against covid: Evidence of covid-19
  signature in cough sounds,'' \emph{arXiv preprint arXiv:2009.08790}, 2020.

\bibitem{kim2018vocal}
B.~Kim, M.~Ghei, B.~Pardo, and Z.~Duan, ``Vocal imitation set: a dataset of
  vocally imitated sound events using the audioset ontology,'' in \emph{DCASE},
  2018.

\bibitem{cartwright2015vocalsketch}
M.~Cartwright and B.~Pardo, ``Vocalsketch: Vocally imitating audio concepts,''
  in \emph{Annual ACM Conference on Human Factors in Computing Systems}, 2015.

\bibitem{tan2019efficientnet}
M.~Tan and Q.~V. Le, ``{EfficientNet}: Rethinking model scaling for
  convolutional neural networks,'' in \emph{ICML}, 2019.

\bibitem{kingma2014adam}
D.~P. Kingma and J.~Ba, ``Adam: A method for stochastic optimization,''
  \emph{arXiv preprint arXiv:1412.6980}, 2014.

\bibitem{park19e_interspeech}
D.~S. Park, W.~Chan, Y.~Zhang, C.-C. Chiu, B.~Zoph, E.~D. Cubuk, and Q.~V. Le,
  ``Specaugment: A simple data augmentation method for automatic speech
  recognition,'' in \emph{Interspeech}, 2019.

\end{thebibliography}

\end{document}